\documentclass[showpacs,preprintnumbers,amsmath,amssymb]{revtex4}
\usepackage{graphicx}
\usepackage{dcolumn}
\usepackage{bm}
\usepackage{epsf}
\newcommand{\p}{\partial}
\newcommand{\omu}{\overline{\mu}}

\newcommand{\occ}{\overline{c}}
\newcommand{\al}{\alpha}

\newcommand{\lms}{\Lambda_{\overline{\mathrm{MS}}}}

\newcommand{\MSbar}{\overline{\mathrm{MS}}}

\newcommand{\Evac}{E_{\textrm{vac}}}
\newcommand{\oo}{\mathcal{O}}
\newcommand{\wsigma}{\widetilde{\sigma}}
\newcommand{\os}{\overline{s}}
\newcommand{\mw}{\mathcal{W}}
\newcommand{\mf}{\mathcal{F}}
\newcommand{\s}{\sigma}
\newcommand{\wj}{\widetilde{J}}
\begin{document}
\preprint{LTH-645}
\title{Off-diagonal mass generation for Yang-Mills theories in the maximal Abelian gauge}
\author{D. Dudal$^a$}
\altaffiliation{Talk given by D. Dudal at ``XXV Encontro Nacional de
F\'{i}sica de Part\'{i}culas e Campos'', Caxambu, Minas Gerais,
Brasil, 24-28 Aug 2004}
 \altaffiliation{Research Assistant of the Fund For Scientific Research-Flanders (Belgium)}
\email{david.dudal@ugent.be}
\author{J.A. Gracey$^b$}
\email{jag@amtp.liv.ac.uk}
\author{V.E.R. Lemes$^c$}
  \email{vitor@dft.if.uerj.br}
\author{M.S. Sarandy$^d$}
\email{msarandy@chem.utoronto.ca}
\author{R.F. Sobreiro$^c$}
 \email{sobreiro@uerj.br}
\author{S.P. Sorella$^c$}
\altaffiliation{Work supported by FAPERJ, Funda{\c c}{\~a}o de
Amparo {\`a} Pesquisa do Estado do Rio de Janeiro, under the program
{\it Cientista do Nosso Estado}, E-26/151.947/2004.}
 \email{sorella@uerj.br}
\author{H. Verschelde$^a$}
 \email{henri.verschelde@ugent.be}
 \affiliation{\vskip 0.1cm $^a$ Ghent University
\\ Department of Mathematical
Physics and Astronomy \\ Krijgslaan 281-S9 \\ B-9000 Gent,
Belgium\\\\ \vskip 0.1cm $^b$ Theoretical Physics Division \\
Department of Mathematical Sciences \\
University of Liverpool \\
P.O. Box 147, Liverpool, L69 3BX, United Kingdom\\\\
\vskip 0.1cm $^c$ UERJ - Universidade do Estado do Rio de
Janeiro\\Rua S\~{a}o Francisco Xavier 524, 20550-013
Maracan\~{a}\\Rio de Janeiro, Brazil\\\\ \vskip 0.1cm $^d$
Chemical Physics Theory Group, Department of Chemistry, University of Toronto, 80 St. George Street, Toronto, Ontario, M5S 3H6, Canada}%
\begin{abstract}
We investigate a dynamical mass generation mechanism for the
off-diagonal gluons and ghosts in $SU(N)$ Yang-Mills theories,
quantized in the maximal Abelian gauge. Such a mass can be seen as
evidence for the Abelian dominance in that gauge. It originates from
the condensation of a mixed gluon-ghost operator of mass dimension
two, which lowers the vacuum energy. We construct an effective
potential for this operator by a combined use of the local composite
operators technique with algebraic renormalization and we discuss
the gauge parameter independence of the results. We also show that
it is possible to connect the vacuum energy, due to the mass
dimension two condensate discussed here, with the non-trivial vacuum
energy originating from the condensate $\left\langle A_\mu^2
\right\rangle$, which has attracted much attention in the Landau
gauge.
\end{abstract}
 \maketitle
\section{\label{sec1}Introduction.}
An unresolved problem of $SU(N)$ Yang-Mills theory is color
confinement. A physical picture that might explain confinement is
based on the mechanism of the dual superconductivity
\cite{scon,'tHooft:1981ht}, according to which the low energy regime
of $QCD$ should be described by an effective Abelian theory in the
presence of magnetic monopoles. These monopoles should condense,
giving rise to the formation of flux tubes which confine the
chromoelectric charges. \\\\Let us provide a very short overview of
the concept of Abelian gauges, which are useful in the search for
magnetic monopoles, a crucial ingredient in the dual
superconductivity picture.
\\\\
\textbf{Abelian gauges.}\\ We recall that $SU(N)$ has a $U(1)^{N-1}$
subgroup, consisting of the diagonal generators. In
\cite{'tHooft:1981ht}, 't Hooft proposed the idea of the Abelian
gauges. Consider a quantity $X(x)$, transforming in the adjoint
representation of $SU(N)$.
\begin{equation}\label{isa1}
    X(x)\rightarrow U(x)X(x)U^+(x)\textrm{ with }U(x)\in SU(N)\;.
\end{equation}
The transformation $U(x)$ which diagonalizes $X(x)$ is the one that
defines the gauge. If $X(x)$ is already diagonal, then clearly
$X(x)$ remains diagonal under the action of the $U(1)^{N-1}$
subgroup. Hence, the gauge is only partially fixed because there is
a residual Abelian gauge freedom.\\\\ In certain space time points
$x_i$, the eigenvalues of $X(x)$ can coincide, so that $U(x_i)$
becomes singular. These possible singularities give rise to the
concept of (Abelian) magnetic monopoles. They have a topological
meaning since $\pi_2\left(SU(N)/U(1)^{N-1}\right)\neq0$ and we refer
to \cite{Kronfeld:1987ri,Kronfeld:1987vd} for all the necessary
details.\\\\\textbf{The dual superconductor as a mechanism behind
confinement.}\\ Let us give a simplified picture of the dual
superconductor to explain the idea. If the $QCD$ vacuum contains
monopoles and if these monopoles condense, there will be a dual
Meissner effect which squeezes the chromoelectric field into a thin
flux tube. This results in a linearly rising potential, $V(r)=\sigma
r$, between static charges, as can be guessed from Gauss' law, $\int
EdS=cte$ or, since the main contribution is coming from the flux
tube, one finds $E\Delta S\approx cte$, hence $V=-\int Edr\approx
cte \times r$
\begin{figure}[t]\label{fig1}
\begin{center}
    \scalebox{1}{\includegraphics{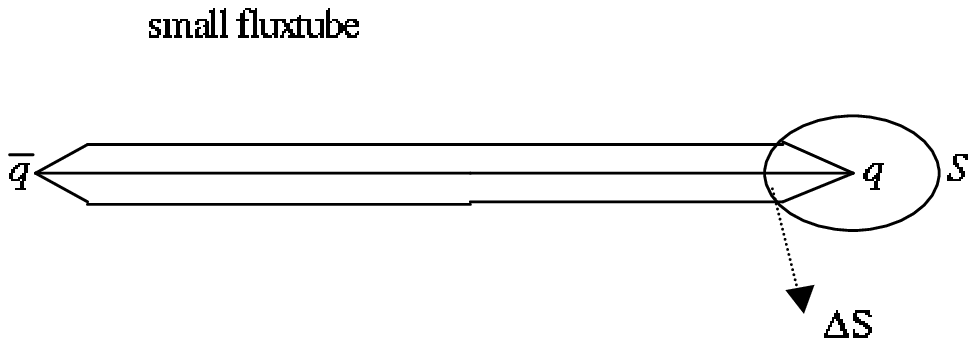}}
    \caption{A chromoelectric flux tube between a static quark-antiquark pair.}
\end{center}
\end{figure}\\\\
\textbf{An example of an Abelian gauge: the maximal Abelian gauge (MAG).}\\
Let $A_{\mu }$ be the Lie algebra valued connection for the gauge
group $SU(N)$, whose generators $T^{A}$, satisfying $\left[
T^{A},T^{B}\right]$~$=$~$f^{ABC}T^{C}$, are chosen to be
antihermitean and to obey the
orthonormality condition $\mathrm{Tr}\left( T^{A}T^{B}\right) =-T_F\delta ^{AB}$%
, with $A,B,C=1,\ldots,\left( N^{2}-1\right) $. In the case of
$SU(N)$, one has $T_F=\frac{1}{2}$. We decompose the gauge field
into its off-diagonal and diagonal parts, namely
\begin{equation}
A_{\mu }=A_{\mu }^{A}T^{A}=A_{\mu }^{a}T^{a}+A_{\mu }^{i}T^{\,i},
\label{conn}
\end{equation}
where  the indices $i$, $j$, $\ldots$ label the $N-1$ generators of
the Cartan subalgebra. The remaining $N(N-1)$ off-diagonal
generators will be labeled by the indices $a$, $b$, $\ldots$. The
field strength decomposes as
\begin{equation}
F_{\mu \nu }=F_{\mu \nu }^{A}T^{A}=F_{\mu \nu }^{a}T^{a}+F_{\mu \nu
}^{i}T^{\,i}\;,  \label{fs}
\end{equation}
with the off-diagonal and diagonal parts given respectively by
\begin{eqnarray}
F_{\mu \nu }^{a} &=&D_{\mu }^{ab}A_{\nu }^{b}-D_{\nu }^{ab}A_{\mu
}^{b}\;+g\,f^{abc}A_{\mu }^{b}A_{\nu }^{c}\;,  \label{fsc} \\
F_{\mu \nu }^{i} &=&\partial _{\mu }A_{\nu }^{i}-\partial _{\nu
}A_{\mu }^{i}+gf^{abi}A_{\mu }^{a}A_{\nu }^{b}\;,  \nonumber
\end{eqnarray}
where the covariant derivative $D_{\mu }^{ab}$ is defined with
respect to the diagonal components $A_{\mu }^{i}$
\begin{equation}
D_{\mu }^{ab}\equiv \partial _{\mu }\delta ^{ab}-gf^{abi}A_{\mu
}^{i}\,\,\,\,\,\,.  \label{cv}
\end{equation}
For the Yang-Mills action one obtains
\begin{equation}
S_{\mathrm{YM}}=-\frac{1}{4}\int d^{4}x\,\left( F_{\mu \nu
}^{a}F^{\mu \nu a}+F_{\mu \nu }^{i}F^{\mu \nu i}\right) \;.
\label{ym}
\end{equation}
The maximal Abelian gauge (MAG), introduced in
\cite{'tHooft:1981ht,Kronfeld:1987vd,Kronfeld:1987ri}, corresponds
to minimizing the functional
\begin{equation}\label{MAGfunctional}
    \mathcal{R}[A]=\int d^4 x\left[A_\mu^a A^{\mu a}\right]
\end{equation}
One checks that $\mathcal{R}[A]$ does exhibit a residual
$U(1)^{N-1}$ invariance.
\\\\The MAG can be recast into a differential form
\begin{equation}\label{MAGdiff}
    D_\mu^{ab}A^{\mu b}=0
\end{equation}
Although we have introduced the MAG here in a functional way, it is
worth mentioning that the MAG does correspond to the diagonalization
of a certain adjoint operator, see e.g. \cite{Amemiya:1998jz}.\\\\
The renormalizability in the continuum of the MAG was proven in
\cite{Min:bx,Fazio:2001rm}, at the cost of introducing a quartic
ghost interaction. The corresponding gauge fixing term turns out to
be \cite{Min:bx,Fazio:2001rm}
\begin{equation}
S_{\mathrm{MAG}}=s\,\int d^{4}x\,\left( \overline{c}^{a}\left(
D_{\mu
}^{ab}A^{b\mu }+\frac{\alpha }{2}b^{a}\right) -\frac{\alpha }{2}gf\,^{abi}%
\overline{c}^{a}\overline{c}^{b}c^{i}-\frac{\alpha }{4}gf\,^{abc}c^{a}%
\overline{c}^{b}\overline{c}^{c}\right) \;,  \label{smn}
\end{equation}
where $\alpha $ is the MAG gauge parameter and $s$ denotes the
nilpotent BRST operator, acting as
\begin{eqnarray}
sA_{\mu }^{a} &=&-\left( D_{\mu }^{ab}c^{b}+gf^{\,abc}A_{\mu
}^{b}c^{c}+gf^{\,abi}A_{\mu }^{b}c^{i}\right) ,\,\,\,\;sA_{\mu
}^{i}=-\left(
\partial _{\mu }c^{i}+gf\,^{iab}A_{\mu }^{a}c^{b}\right) \;,  \nonumber \\
sc^{a} &=&gf\,^{abi}c^{b}c^{i}+\frac{g}{2}f\,^{abc}c^{b}c^{c},\,\,\,\,\,\,\,%
\,\,\,\,\,\,\,\,\,\,\,\,\,\,\,\,\,\,\,\,\,\,\,\,\,\,\,\,\,\,\,\,\,\,\,sc^{i}=%
\frac{g}{2}\,f\,^{iab}c^{a}c^{b},  \nonumber \\
s\overline{c}^{a}
&=&b^{a}\;,\,\,\,\,\,\,\,\,\,\,\,\,\,\,\,\,\,\,\,\,\,\,\,\,\,\,\,\,\,\,\,\,%
\,\,\,\,\,\,\,\,\,\,\,\,\,\,\,\,\,\,\,\,\,\,\,\,\,\,\,\,\,\,\,\,\,\,\,\,\,\,%
\,\,\,\,\,\,\,\,\,\,\,\,\,\,\,\,\,\,\,\,\,\,s\overline{c}^{i}=b^{i}\;,  \nonumber \\
sb^{a}
&=&0\;,\,\,\,\,\,\,\,\,\,\,\,\,\,\,\,\,\,\,\,\,\,\,\,\,\,\,\,\,\,\,\,\,\,\,%
\,\,\,\,\,\,\,\,\,\,\,\,\,\,\,\,\,\,\,\,\,\,\,\,\,\,\,\,\,\,\,\,\,\,\,\,\,\,%
\,\,\,\,\,\,\,\,\,\,\,\,\,\,\,\,\,\,\,\,\,\,sb^{i}=0\;. \label{brst}
\end{eqnarray}
Here $c^{a},c^{i}$ are the off-diagonal and the diagonal components
of the Faddeev-Popov ghost field, while $\overline{c}^{a},b^{a}$ are
the off-diagonal antighost and Lagrange multiplier. We also observe
that the BRST\ transformations $\left( \ref{brst}\right) $ have been
obtained by their standard form upon projection on the off-diagonal
and diagonal components of the fields.  We remark that the MAG
(\ref{smn}) can be written in the form
\begin{equation}
S_{\mathrm{MAG}}=s\os\int d^{4}x\,\left(\frac{1}{2}A_\mu^a A^{\mu
a}-\frac{\al}{2}c^a\occ^a\right) \;, \label{smn2}
\end{equation}
with $\os$  being the nilpotent anti-BRST transformation, acting as
\begin{eqnarray}
\os A_{\mu }^{a} &=&-\left( D_{\mu }^{ab}\occ^{b}+gf^{\,abc}A_{\mu
}^{b}\occ^{c}+gf^{\,abi}A_{\mu }^{b}\occ^{i}\right)
,\,\,\,\,\,\,\,\;\os A_{\mu }^{i}=-\left(
\partial _{\mu }\occ^{i}+gf\,^{iab}A_{\mu }^{a}\occ^{b}\right) \;,  \nonumber \\
\os\occ^{a} &=&gf^{abi}\occ^{b}\occ^{i}+\frac{g}{2}f^{abc}\occ^{b}\occ^{c},\,\,\,\,\,\,\,%
\,\,\,\,\,\,\,\,\,\,\,\,\,\,\,\,\,\,\,\,\,\,\,\,\,\,\,\,\,\,\,\,\,\,\,\,\,\,\,\,\,\,\,\os\occ^{i}=%
\frac{g}{2}\,f\,^{iab}\occ^{a}\occ^{b},  \nonumber \\
\os c^{a}
&=&-b^{a}+gf^{abc}c^b\occ^c+gf^{abi}c^b\occ^i+gf^{abi}\occ^bc^i\;,\,\,\,\,\os c^{i}=-b^{i}+gf^{ibc}c^b\occ^c\;,  \nonumber \\
\os b^{a} &=&-gf^{abc}b^b\occ^c-gf^{abi}b^b\occ^i+gf^{abi}\occ^b
b^i\;\;\;\;\;\;\;\;\;\;\;\;\;\os b^i=-gf^{ibc}b^b\occ^c\;.
\label{brst2}
\end{eqnarray}
It can be checked that $s$ and $\os$ anticommute. \\\\Expression
$\left( \ref{smn}\right) $ is easily worked out and yields
\begin{eqnarray}
S_{\mathrm{MAG}} &=&\int d^{4}x\left( b^{a}\left( D_{\mu }^{ab}A^{\mu b}+%
\frac{\alpha }{2}b^{a}\right) +\overline{c}^{a}D_{\mu }^{ab}D^{\mu bc}c^{c}+g%
\overline{c}^{a}f^{abi}\left( D_{\mu }^{bc}A^{\mu c}\right)
c^{i}+g\overline{c}^{a}D_{\mu }^{ab}\left( f^{bcd}A^{\mu c
}c^{d}\right) \right.\nonumber\\&-&\alpha
gf^{abi}b^{a}\overline{c}^{b}c^{i}-\left.g^{2}f^{abi}f^{cdi}\overline{c}^{a}c^{d}A_{\mu
}^{b}A^{\mu c}-\frac{\alpha }{2}gf^{abc}b^{a}\overline{c}%
^{b}c^{c}-\frac{\alpha }{4}g^{2}f^{abi}f^{cdi}\overline{c}^{a}\overline{c}%
^{b}c^{c}c^{d}\right.\nonumber\\&-&\left.\frac{\alpha }{4}g^{2}f^{abc}f^{adi}\overline{c}^{b}\overline{c}%
^{c}c^{d}c^{i}  -\frac{\alpha }{8}g^{2}f^{abc}f^{ade}%
\overline{c}^{b}\overline{c}^{c}c^{d}c^{e}\right)\;. \label{smne}
\end{eqnarray}
We note that $\al=0$ does in fact correspond to the ``real'' MAG
condition, given by eq.(\ref{MAGdiff}). However, one cannot set
$\al=0$ from the beginning since this would lead to a
nonrenormalizable gauge. Some of the terms proportional to $\al$
would reappear due to radiative corrections, even if $\al=0$. See,
for example, \cite{Kondo:1997pc}. For our purposes, this means that
we have to keep $\al$ general  throughout and leave to the end the
analysis of the limit $\al\rightarrow 0$, to recover condition
(\ref{MAGdiff}).\\\\ In order to have a complete quantization of the
theory, one has to fix the residual Abelian gauge freedom by means
of a suitable further gauge condition on the diagonal components
$A_{\mu }^{i}$ of the gauge field.  A common choice for the Abelian
gauge fixing, also adopted in the lattice papers
\cite{Amemiya:1998jz,Bornyakov:2003ee}, is the Landau gauge, given
by
\begin{equation}
S_{\mathrm{diag}}=s\,\int d^{4}x\,\;\overline{c}^{i}\partial _{\mu
}A^{\mu i
}\;=\int d^{4}x\,\;\left( b^{i}\partial _{\mu }A^{\mu i}+\overline{c}%
^{i}\partial ^{\mu }\left( \partial _{\mu }c^{i}+gf\,^{iab}A_{\mu
}^{a}c^{b}\right) \right) \;,  \label{abgf}
\end{equation}
where $\overline{c}^{i},b^{i}$ are the diagonal antighost and
Lagrange multiplier.\\\\
\textbf{Abelian dominance.}\\
According to the concept of Abelian dominance, the low energy regime
of $QCD$ can be expressed solely in terms of Abelian degrees of
freedom \cite{Ezawa:bf}. Lattice confirmations of the Abelian
dominance can be found in \cite{Suzuki:1989gp,Hioki:1991ai}. To our
knowledge, there is no analytic proof  of the Abelian dominance.
Nevertheless, an argument that can be interpreted  as evidence of
it, is the fact that the off-diagonal gluons would attain a
dynamical mass. At energies below the scale set by this mass, the
off-diagonal gluons should decouple, and in this way one should end
up with an Abelian theory at low energies.\\\\A lattice study of
such an off-diagonal gluon mass reported a value of approximately
$1.2$GeV \cite{Amemiya:1998jz}. More recently, the off-diagonal
gluon propagator was investigated numerically in
\cite{Bornyakov:2003ee}, reporting a similar result.\\\\
There have been  several efforts to give an analytic description of
the mechanism responsible for the dynamical generation of the
off-diagonal gluon mass. In \cite{Schaden:1999ew,Kondo:2000ey}, a
certain ghost condensate was used to construct an effective,
off-diagonal mass. However, in \cite{Dudal:2002xe} it was shown that
the obtained mass was a tachyonic one, a fact confirmed later in
\cite{Sawayanagi:2003dc}. Another condensation, namely that of the
mixed gluon-ghost operator $(\frac{1}{2}A_\mu^a A^{\mu a}+\al\occ^a
c^a)$ \footnote{The index $a$ runs only over the $N(N-1)$
off-diagonal generators.}, that could be responsible for the
off-diagonal mass, was proposed in \cite{Kondo:2001nq}. That this
operator should condense can be expected on the basis of a close
analogy existing between the MAG and the renormalizable nonlinear
Curci-Ferrari gauge \cite{Curci:bt,Curci:1976ar}. In fact, it turns
out that the mixed gluon-ghost operator can be introduced also in
the Curci-Ferrari gauge. A detailed analysis of its condensation and
of the ensuing dynamical mass generation can be found in
\cite{Dudal:2003pe,Dudal:2003gu}.
\\\\Here, we shall report on the results of \cite{Dudal:2004rx}. It was
investigated explicitly if the mass dimension two operator
$(\frac{1}{2}A_\mu^a A^{\mu a}+\al\occ^a c^a)$ condenses, so that a
dynamical off-diagonal mass is generated in the MAG. The pathway we
intend to follow is based on previous research in this direction in
other gauges. In \cite{Verschelde:2001ia}, the local composite
operator (LCO) technique was used to construct a renormalizable
effective potential for the operator $A_\mu^A A^{\mu A}$ in the
Landau gauge. As a consequence of $\left\langle A_\mu^A A^{\mu
A}\right\rangle\neq0$, a dynamical mass parameter is generated
\cite{Verschelde:2001ia}. The condensate $\left\langle A_\mu^A
A^{\mu A}\right\rangle$ has attracted attention from theoretical
\cite{Gubarev:2000eu,Gubarev:2000nz} as well as from the lattice
side \cite{Boucaud:2001st}. It was shown by means of the algebraic
renormalization technique \cite{Piguet:1995er} that the LCO
formalism  for the condensate $\left\langle A_\mu^A A^{\mu
A}\right\rangle$ is renormalizable to all orders of perturbation
theory \cite{Dudal:2002pq}.  The same formalism was successfully
employed to study the condensation of $(\frac{1}{2}A_\mu^A A^{\mu
A}+\al\occ^A c^A)$ in the Curci-Ferrari gauge
\cite{Dudal:2003pe,Dudal:2003gu}. We would like to note that the
Landau gauge corresponds to $\al=0$. Later on, the condensation of
$A_\mu^A A^{\mu A}$ was confirmed in the linear covariant gauges
\cite{Dudal:2003np,Dudal:2003by}, which also possess the Landau
gauge as a special case. It was proven formally that the vacuum
energy does not depend on the gauge parameter in these gauges. As
such, the linear, Curci-Ferrari and Landau gauges are all connected
to each other. We managed to connect also the MAG with the Landau
gauge, and as such with the linear and Curci-Ferrari gauges
\cite{Dudal:2004rx}.
\section{Renormalizability of $SU(N)$ Yang-Mills theories in the MAG in the presence of the
local composite operator $(\frac{1}{2}A_\mu^a A^{\mu a}+\al\occ^a
c^a)$ .} To prove the renormalizability to all orders of
perturbation theory, we shall rely on the algebraic renormalization
formalism \cite{Piguet:1995er}. In order to write down a suitable
set of Ward identities, we first introduce external fields
$\Omega^{\mu i}$, $\Omega^{\mu a}$, $L^{i}$, $L^{a}$ coupled to the
BRST\ nonlinear variations of the fields, namely
\begin{eqnarray}
S_{\mathrm{ext}} &=&\int d^{4}x\left( -\Omega ^{\mu a}\left(
D_{\mu }^{ab}c^{b}+gf^{abc}A_{\mu }^{b}c^{c}+gf^{abi}A_{\mu
}^{b}c^{i}\right)  -\Omega^{\mu i}\left( \partial _{\mu
}c^{i}+gf^{iab}A_{\mu }^{a}c^{b}\right)\right.\nonumber\\ &+&\left.L^{a}\left( gf^{abi}c^{b}c^{i}+%
\frac{g}{2}f^{abc}c^{b}c^{c}\right)
+L^{i}\frac{g}{2}\,f^{iab}c^{a}c^{b}\right) \;,   \label{sexr}
\end{eqnarray}
with
\begin{eqnarray}
s\Omega^{\mu a} &=&s\Omega^{\mu i}=0\;,  \label{ss} \\
sL^{a} &=&sL^{i}=0\;.  \nonumber
\end{eqnarray}
Moreover, in order to discuss the renormalizability of the
gluon-ghost operator
\begin{equation}
\mathcal{O}_{\mathrm{MAG}}= \frac{1}{2}A_{\mu}^{a}A^{\mu a}+\alpha \overline{c}%
^{a}c^{a} \;,  \label{ggop}
\end{equation}
we introduce it in the starting action by means of a BRST\ doublet
of external sources $\left( J,\lambda \right)$
\begin{equation}
s\lambda =J\;,\;\;\;\;sJ=0\;,  \label{jl}
\end{equation}
so that
\begin{eqnarray}
S_{\mathrm{LCO}} &=&s\int d^{4}x\;\left( \lambda \left(
\frac{1}{2}A_{\mu
}^{a}A^{\mu a}+\alpha \overline{c}^{a}c^{a}\right) +\zeta \frac{\lambda J}{2%
}\right) \;  \label{slco} \\
&=&\int d^{4}x\;\left( J\left( \frac{1}{2}A_\mu^a A^{\mu a}+\alpha
\overline{c}^{a}c^{a}\right) +\zeta \frac{J^{2}}{2}-\alpha \lambda
b^{a}c^{a}\right.   \nonumber \\
&+&\left. \lambda A^{\mu a}\left( D_{\mu
}^{ab}c^{b}+gf^{abi}A_{\mu }^{b}c^{i}\right) +\alpha \lambda
\overline{c}^{a}\left(
gf\,^{abi}c^{b}c^{i}+\frac{g}{2}f\,^{abc}c^{b}c^{c}\right) \right)
\;, \nonumber
\end{eqnarray}
where $\zeta $ is the LCO parameter accounting for the divergences
present in the vacuum correlator
$\left\langle\mathcal{O}_{\mathrm{MAG}}(x)\mathcal{O}_{\mathrm{MAG}}(y)
\right\rangle $, which are proportional to $J^{2}$. Therefore, the
complete action
\begin{equation}
\Sigma =S_{\mathrm{YM}}+S_{\mathrm{MAG}}+S_{\mathrm{diag}}+S_{\mathrm{ext}%
}+S_{\mathrm{LCO}}\;,  \label{ca}
\end{equation}
is BRST invariant
\begin{equation}
s\Sigma =0\;.  \label{inv}
\end{equation}
As noticed in \cite{Kondo:2001nq,Kondo:2001tm}, the gluon-ghost mass
operator defined in eq.(\ref{ggop}) is BRST invariant on-shell. We
have written down in \cite{Dudal:2004rx} all the Ward identities,
which are sufficient to prove that the most general local
counterterm, compatible with the symmetries of the model, can always
be reabsorbed by means of multiplicative renormalization. As an
interesting by-product, we have been able to establish a relation
between the anomalous dimension of the gluon-ghost operator
$\mathcal{O}_{\mathrm{MAG}}$ and other, more elementary,
renormalization group functions. Explicitly, it holds to all orders
of perturbation theory that
\begin{equation}
\gamma_{\mathcal{O}_{\mathrm{MAG}}}(g^2)=-2\left(\frac{\beta(g^2)
}{2g^2}-\gamma _{c^{i}}(g^2)\right), \label{go}
\end{equation}
where $\beta(g^2)=\mu\frac{\p g^2}{\p\mu}$ and $\gamma_{c^i}(g^2)$
denotes the anomalous dimension of the diagonal ghost field.
\subsection{The effective potential via the LCO method.}
We present here the main steps in the construction of the effective
potential for a local composite operator. A more detailed account of
the LCO formalism can be found in
\cite{Verschelde:jj,Knecht:2001cc}.\\\\To obtain the effective
potential for the condensate
$\left\langle\mathcal{O}_{\mathrm{MAG}}\right\rangle$, we set the
sources $\Omega_\mu^i$, $\Omega_\mu^a$, $L^a$, $L^i$ and $\lambda$
to zero and consider the renormalized generating functional
\begin{eqnarray}
\label{d1}    \exp(-i\mw(J))&=&\int [D\varphi]\exp iS(J)\;,\nonumber\\
\label{d2}
    S(J)&=&S_{\mathrm{YM}}+S_{\mathrm{MAG}}+S_{\mathrm{diag}}+S_{\mathrm{count}}+\int d^{4}x
\left(Z_JJ\left(\frac{1}{2}\widetilde{Z}_A A_\mu^a A^{\mu
a}+Z_\al\widetilde{Z}_c\al \occ^a
c^a\right)+(\zeta+\delta\zeta)\frac{J^2}{2}\right)\;,
\end{eqnarray}
where $\varphi$ denotes the relevant fields and
$S_{\mathrm{count}}$ is the usual counterterm contribution, i.e.
the part without the composite operator. The quantity
$\delta\zeta$ is the counterterm accounting for the divergences
proportional to $J^2$. Using dimensional regularization throughout
with the convention that $d=4-\varepsilon$, one has the following
identification
\begin{equation}\label{rge5}
    \zeta_0J_0^2=\mu^{-\varepsilon}(\zeta+\delta\zeta)J^2\;.
\end{equation}
The  functional $\mw(J)$ obeys the renormalization group equation
(RGE)
\begin{equation}\label{rge3}
    \left(\mu\frac{\p}{\p\mu}+\beta(g^2)\frac{\p}{\p
g^2}+\al\gamma_\al(g^2)\frac{\p}{\p\al}-\gamma_{\mathcal{O}_{\mathrm{MAG}}}(g^2)\int
d^4x J\frac{\delta}{\delta
    J}+\eta(g^2,\zeta)\frac{\p}{\p\zeta}\right)\mw(J)=0\;,
\end{equation}
where
\begin{eqnarray}\label{rge4}
    \gamma_{\al}(g^2)&=&\mu\frac{\p}{\p\mu}\ln\al\;,\nonumber\\
\eta(g^2,\zeta)&=&\mu\frac{\p}{\p\mu}\zeta\;.
\end{eqnarray}
>From eq.(\ref{rge5}), one finds
\begin{equation}\label{rge6}
\eta(g^2,\zeta)=2\gamma_{\mathcal{O}_{\mathrm{MAG}}}(g^2)\zeta+\delta
(g^2,\al)\;,
\end{equation}
with
\begin{equation}  \label{rge7}
\delta(g^{2},\alpha)=\left(\varepsilon+2\gamma_{\mathcal{O}_{\mathrm{MAG}}}(g^{2})-\beta
(g^{2})\frac{\partial }{\partial g^{2}}%
-\alpha\gamma_{\alpha}(g^{2})\frac{\partial}{\partial\alpha}%
\right)\delta\zeta\;.
\end{equation}
Up to now, the LCO parameter $\zeta$ is still an arbitrary
coupling. As explained in \cite{Verschelde:jj,Knecht:2001cc},
simply setting $\zeta=0$ would give rise to an inhomogeneous RGE
for $\mw(J)$
\begin{equation}  \label{rge8}
    \left(\mu\frac{\p}{\p\mu}+\beta(g^2)\frac{\p}{\p
g^2}+\al\gamma_\al(g^2)\frac{\p}{\p\al}-\gamma_{\mathcal{O}_{\mathrm{MAG}}}(g^2)\int
d^4x J\frac{\delta}{\delta
    J}\right)\mw(J)=\delta(g^2,\al)\int d^4x \frac{J^2}{2}\;,
\end{equation}
and a non-linear RGE for the associated effective action $\Gamma$
for the composite operator $\mathcal{O}_{\mathrm{MAG}}$.
Furthermore, multiplicative renormalizability is lost and by
varying the value of $\delta\zeta$, minima of the effective action
can change into maxima or can get lost. However, $\zeta$ can be
made such a function of $g^2$ and $\al$ so that, if $g^2$ runs
according to $\beta(g^2)$ and $\al$ according to
$\gamma_\al(g^2)$, $\zeta(g^2,\al)$ will run according to its RGE
(\ref{rge6}). This is accomplished by setting $\zeta$ equal to the
solution of the differential equation
\begin{equation}\label{rge9}
\left(\beta(g^2)\frac{\partial}{\partial
g^2}+\al\gamma_\al(g^2,\al)\frac{\partial}{\partial\al}\right)\zeta(g^2,
\al)=2\gamma_{\mathcal{O}_{\mathrm{MAG}}}(g^2)\zeta(g^2,\alpha)+\delta(g^2,\al)\;.
\end{equation}
Doing so, $\mw(J)$ obeys the homogeneous renormalization group
equation
\begin{equation}  \label{rge10}
    \left(\mu\frac{\p}{\p\mu}+\beta(g^2)\frac{\p}{\p
g^2}+\al\gamma_\al(g^2)\frac{\p}{\p\al}-\gamma_{\mathcal{O}_{\mathrm{MAG}}}(g^2)\int
d^4x J\frac{\delta}{\delta J}\right)\mw(J)=0\;.
\end{equation}
To lighten the notation, we will drop the renormalization factors
from now on. One will notice that there are terms quadratic in the
source $J$ present in $\mathcal{W}(J)$, obscuring the usual energy
interpretation. This can be cured  by removing the terms
proportional to $J^2$ in the action to get a generating functional
that is linear in the source, a goal easily achieved by inserting
the following unity,
\begin{equation}  \label{rge11}
1=\frac{1}{N}\int [D\sigma]\exp\left[i\int
d^{4}x\left(-\frac{1}{2\zeta}\left(%
\frac{\sigma}{g}- \mathcal{O}_{\mathrm{MAG}}-\zeta
J\right)^{2}\right)\right]\;,
\end{equation}
with $N$ the appropriate normalization factor, in eq.(\ref{d1}) to
arrive at the Lagrangian
\begin{eqnarray}
\mathcal{L}(A_\mu,\sigma)=-\frac{1}{4}F_{\mu\nu}^{a}F^{\mu\nu
a}-\frac{1}{4}F_{\mu\nu}^{i}F^{\mu\nu
i}+\mathcal{L}_{\textrm{MAG}}+\mathcal{L}_{\textrm{diag}}-\frac{\sigma^2}{2g^2\zeta}
+\frac{1}{g^2\zeta}g\sigma\mathcal{O}_{\mathrm{MAG}}-\frac{1}{2\zeta}\left(\mathcal{O}_{\mathrm{MAG}}\right)^2\;,
\label{rge12}
\end{eqnarray}
while
\begin{eqnarray}
\label{rge13}\exp(-i\mw(J))&=&\int [D\varphi]\exp iS_\sigma(J)\;,\\
\label{rge13bis}S_\sigma(J)&=&\int
d^4x\left(\mathcal{L}(A_\mu,\sigma)+J\frac{\sigma}{g}\right)\;.
\end{eqnarray}
>From eqs.(\ref{d1}) and (\ref{rge13}), one has the following simple
relation
\begin{equation}\label{rge14}
    \left.\frac{\delta\mw(J)}{\delta
J}\right|_{J=0}=-\left\langle\mathcal{O}_{\mathrm{MAG}}\right\rangle=-\left\langle
\frac{\sigma}{g}\right\rangle\;,
\end{equation}
meaning that the condensate $\left\langle
\mathcal{O}_{\mathrm{MAG}}\right\rangle$ is directly related to the
expectation value of the field $\s$, evaluated with the action
$S_\s=\int d^4x\mathcal{L}(A_\mu,\s)$. As it is obvious from
eq.(\ref{rge12}), $\left\langle\sigma\right\rangle\neq0$ is
sufficient to have a tree level dynamical mass for the off-diagonal
fields. At lowest order (i.e. tree level), one finds
\begin{eqnarray}
m_{\mathrm{gluon}}^{\mathrm{off-diag.}}&=&\sqrt{\frac{g\sigma}{\zeta_0}}\;,\;\;\;\;\;\;\;\;\;\;
m_{\mathrm{ghost}}^{\mathrm{off-diag.}}=\sqrt{\al
\frac{g\sigma}{\zeta_0}}\;.
\end{eqnarray}
Meanwhile, the diagonal degrees of freedom remain massless.
\section{Gauge parameter independence of the vacuum energy.}
We begin this section with a few remarks on the determination of
$\zeta(g^2,\al)$. From explicit calculations in perturbation
theory, it will become clear \footnote{See section V.} that the
RGE functions showing up in the differential equation (\ref{rge9})
look like
\begin{eqnarray}\label{ga1}
    \beta(g^2)&=&-\varepsilon
    g^2-2\left(\beta_0g^2+\beta_1g^2+\cdots\right)\;,\nonumber\\
    \gamma_{\mathcal{O}_{\mathrm{MAG}}}(g^2)&=&\gamma_0(\al)g^2+\gamma_1(\al)g^4+\cdots\;,\nonumber\\
    \gamma_{\al}(g^2)&=&a_0(\al)g^2+a_1(\al)g^4+\cdots\;,\nonumber\\
    \delta(g^2,\al)&=&\delta_0(\al)+\delta_1(\al)g^2+\cdots\;.
\end{eqnarray}
As such, eq.(\ref{rge9}) can be solved  by expanding
$\zeta(g^2,\al)$ in a Laurent series in $g^2$,
\begin{equation}\label{ga2}
    \zeta(g^2,\al)=\frac{\zeta_0(\al)}{g^2}+\zeta_1(\al)+\zeta_2(\al)g^2+\cdots\;.
\end{equation}
More precisely, for the first coefficients $\zeta_0$, $\zeta_1$ of
the expression (\ref{ga2}), one obtains
\begin{eqnarray}
\label{rge21a}2\beta_0\zeta_0+\al
a_0\frac{\partial\zeta_0}{\partial\al}&=&2\gamma_0\zeta_0+\delta_0\;,\nonumber\\
\label{rge21b}2\beta_1\zeta_0+\al
a_0\frac{\partial\zeta_1}{\partial\al}+\al
a_1\frac{\partial\zeta_0}{\partial\al}&=&2\gamma_0\zeta_1+2\gamma_1\zeta_0
+\delta_1\;.
\end{eqnarray}
Notice that, in order to construct the $n$-loop effective potential,
knowledge of the $(n+1)$-loop RGE functions is needed. \\\\The
effective potential calculated with the Lagrangian (\ref{rge12})
will explicitly depend on the gauge parameter $\al$. The question
arises concerning the vacuum energy $\Evac$, (i.e. the effective
potential evaluated at its minimum); will it be independent of the
choice of $\al$? Also, as it can be seen from the equations
(\ref{rge21a}), each $\zeta_i(\al)$ is determined through a first
order differential equation in $\al$. Firstly, one has to solve for
$\zeta_0(\al)$. This will introduce one arbitrary integration
constant $C_0$. Using the obtained value for $\zeta_0(\al)$, one can
consequently solve the first order differential equation for
$\zeta_1(\al)$. This will introduce a second integration constant
$C_1$, etc. In principle, it is possible that these arbitrary
constants influence the vacuum energy,  which would represent an
unpleasant feature. Notice that the differential equations in $\al$
for the $\zeta_i$ are due to the running of $\al$ in
eq.(\ref{rge9}), encoded in the renormalization group function
$\gamma_\al(g^2)$. Assume that we would have already shown that
$\Evac$ does not depend on the choice of $\al$. If we then set
$\al=\al^*$, with $\al^*$ a fixed point of the RGE for $\al$ at the
considered order of perturbation theory, then equation (\ref{rge9})
determining $\zeta$ simplifies to
\begin{equation}\label{rge9bis}
\beta(g^2)\frac{\partial}{\partial g^2}\zeta(g^2,
\al^*)=2\gamma_{\mathcal{O}_{\mathrm{MAG}}}(g^2)\zeta(g^2,\alpha^*)+\delta(g^2,\al^*)\;,
\end{equation}
since
\begin{equation}\label{rge9tris}
    \left.\gamma_\al(g^2)\al\right|_{\al=\al^*}=0\;.
\end{equation}
This will lead to simple algebraic equations for the
$\zeta_i(\al^*)$. Hence, no integration constants will enter the
final result for the vacuum energy for $\al=\al^*$, and since
$\Evac$ does not depend on $\al$, $\Evac$ will never depend on the
integration constants, even when calculated for a general $\al$.
Hence, we can put them equal to zero from the beginning for
simplicity.\\\\Summarizing, two questions remain. Firstly, we should
prove that the value of $\al$ will not influence the obtained value
for $\Evac$. Secondly, we should show that there exists a fixed
point $\al^*$. We postpone the discussion concerning the second
question to the next section, giving a positive answer to the first
one. In order to do so,  let us reconsider the generating functional
(\ref{rge13}). We have the following identification, ignoring the
overall normalization factors
\begin{eqnarray}\label{rge34}
    \exp(-i\mw(J))=\int [D\varphi]\exp iS_\sigma(J)=\frac{1}{N}\int
    [D\varphi D\sigma]\exp i\left[S(J)+\int
d^{4}x\left(-\frac{1}{2\zeta}\left(%
\frac{\sigma}{g}-\mathcal{O}_{\mathrm{MAG}}-\zeta
J\right)^{2}\right)\right]\;,
\end{eqnarray}
where $S(J)$ and $S_\sigma(J)$ are given respectively by
eq.(\ref{d2}), and eq.(\ref{rge13bis}). Obviously,
\begin{equation}
\frac{d}{d\alpha }\frac{1}{N}\int [D\sigma ]\exp \left[ i\int
d^{4}x\left( -%
\frac{1}{2\zeta }\left( \frac{\sigma
}{g}-\mathcal{O}_{\mathrm{MAG}}-\zeta J\right) ^{2}\right) \right]
=\frac{d}{d\alpha }1=0\;, \label{rge35}
\end{equation}
so that
\begin{equation}
\frac{d\mathcal{W}(J)}{d\alpha }=-\left.\left\langle s\int
d^{4}x\os\left(\frac{1}{2}c^a\occ^a\right)
 \right\rangle\right|_{J=0}+\textrm{terms}\propto J\;, \label{rge36}
\end{equation}
which follows directly from
\begin{eqnarray}
\frac{dS(J)}{d\alpha } =s\os\int
d^4x\left(\frac{1}{2}c^a\occ^a\right)+\textrm{terms}\propto J\;.
\label{rge37}
\end{eqnarray}
We see that the first term in the right hand side of (\ref{rge37})
is an exact BRST variation. As such, its vacuum expectation value
vanishes. This is the usual argument to prove the gauge parameter
independence in the BRST framework \cite{Piguet:1995er}. Note that
no local operator  $\mathcal{\hat O}$, with $s\mathcal{\hat
O}=\mathcal{O}_{\mathrm{MAG}}$, exists. Furthermore, extending the
action of the BRST transformation on the $\sigma$-field by
\begin{equation}\label{rge38}
    s\s=gs\mathcal{O}_{\mathrm{MAG}}=-A^{\mu a}D_\mu^{ab}c^b+\al b^a c^a-\al gf^{abi}\occ^a\occ^bc^i-\frac{\al}{2}gf^{abc}\occ^a c^b c^c
\end{equation}
one can easily check that
\begin{equation}\label{rge39}
    s\int d^{4}x\mathcal{L}(A_\mu,\s)=0\;,
\end{equation}
so that we have a BRST invariant $\sigma$-action. Thus, when we
consider the vacuum, corresponding to $J=0$, only the BRST exact
term in eq.(\ref{rge36}) survives. The effective action $\Gamma$ is
related to $\mathcal{W}(J)$ through a Legendre transformation
$\Gamma \left( \frac{\sigma }{g}\right) =-\mathcal{W}(J)-\int
d^{4}yJ(y)\frac{%
\sigma (y)}{g}$, while the effective potential $V(\sigma )$ is
defined as
\begin{equation}
-V(\sigma )\int d^{4}x=\Gamma \left( \frac{\sigma }{g}\right)\;.
\label{rge41}
\end{equation}
If $\sigma _{\mathrm{min}}$ is the solution of $\frac{dV(\sigma
)}{d\sigma } =0$, then it follows from $\frac{\delta }{\delta \left(
\frac{\sigma }{g}\right) }\Gamma=-J$, that
\begin{equation}
\sigma =\sigma _{\mathrm{min}}\Rightarrow J=0\;,  \label{rge43}
\end{equation}
and hence,
\begin{equation}
\left. \frac{d}{d\alpha }V(\sigma )\right| _{\sigma =\sigma
_{\mathrm{min}}}\int d^{4}x=\left. \frac{d}{d\alpha
}\mathcal{W}(J)\right| _{J=0}\;,  \label{rge44}
\end{equation}
or, due to eq.(\ref{rge36}),
\begin{equation}
\left. \frac{d}{d\alpha }V(\sigma )\right| _{\sigma =\sigma
_{\mathrm{min}}}=0\;. \label{rge45}
\end{equation}
We conclude that the vacuum energy $\Evac$ should be independent
from the gauge parameter $\al$.
\\\\A completely analogous derivation was performed in the case of the
linear gauge \cite{Dudal:2003by}. Nevertheless, in spite of the
previous argument, explicit results in that case showed that $\Evac$
did depend on $\al$.  In \cite{Dudal:2003by} it was argued that this
apparent disagreement was due to a mixing of different orders of
perturbation theory. We proposed a modification of the LCO formalism
suitable circumventing this problem and obtaining a well defined
gauge independent vacuum energy $\Evac$, without the need of working
at infinite order \cite{Dudal:2003by}. Instead of the action
(\ref{d2}), let us consider the following action
\begin{eqnarray}
\widetilde{S}(\wj)&=&S_{\mathrm{YM}}+S_{\mathrm{MAG}}+S_{\mathrm{diag}}+\int
d^{4}x\left[
\widetilde{J}\mf(g^2,\al)\mathcal{O}_{\mathrm{MAG}} +\frac{\zeta }{2}%
\mf^2(g^2,\al)\widetilde{J}^{2}\right]\;, \label{rge53}
\end{eqnarray}
where, for the moment, $\mf(g^2,\al)$ is an arbitrary function of
$\al$ of the form
\begin{equation}\label{rge54}
    \mf(g^2,\al)=1+f_{0}(\al)g^2+f_1(\al)g^4+O(g^6)\;,
\end{equation}
and $\wj$ is now the source. The generating functional becomes
\begin{equation}\label{rge55}
    \exp(-i\mathcal{\widetilde{W}}(\wj))=\int[D\phi]\exp
i\widetilde{S}(\wj)\;.
\end{equation}
Taking the functional derivative of $\mathcal{\widetilde{W}}(\wj)$
with respect to $\wj$, we obtain
\begin{equation}
\left. \frac{\delta \mathcal{\widetilde{W}}(\wj)}{\delta
\wj}\right|
_{\wj=0}=-\mf(g^2,\al)\left\langle\mathcal{O}_{\mathrm{MAG}}\right\rangle\;.
\label{rge56}
\end{equation}
Once more, we insert unity via
\begin{equation}  \label{rge57}
1=\frac{1}{N}\int [D\wsigma]\exp\left[i\int
d^{4}x\left(-\frac{1}{2\zeta}\left(%
\frac{\wsigma}{g\mf(g^2,\al)}-\mathcal{O}_{\mathrm{MAG}}-\zeta
\wj\mf(g^2,\al) \right)^{2}\right)\right]\;,
\end{equation}
to arrive at the following Lagrangian
\begin{eqnarray}\label{rge58}
\mathcal{\widetilde{L}}(A_\mu,\wsigma)=-\frac{1}{4}F_{\mu\nu}^{a}F^{\mu\nu
a}-\frac{1}{4}F_{\mu\nu}^{i}F^{\mu\nu i}
+\mathcal{L}_{\mathrm{MAG}}+\mathcal{L}_{\mathrm{diag}}-\frac{\wsigma^2}{2g^2\mf^2(g^2,\al)\zeta}
    +\frac{1}{g^2\mf(g^2,\al)\zeta}g\wsigma
\mathcal{O}_{\mathrm{MAG}}-\frac{1}{2\zeta}\left(\mathcal{O}_{\mathrm{MAG}}
\right)^2 .
\end{eqnarray}
>From the generating functional
\begin{eqnarray}
    \label{rge59}\exp(-i\mathcal{\widetilde{W}}(\wj))&=&\int[D\phi]\exp iS_{\wsigma}(\wj)\;,\\
\label{rge59bis}S_{\wsigma}(\wj)&=&\int
d^4x\left(\mathcal{L}(A_\mu,\wsigma)+\wj\frac{\wsigma}{g}\right)\;.
\end{eqnarray}
it follows that
\begin{eqnarray}
\left. \frac{\delta \mathcal{\widetilde{W}}(\wj)}{\delta
\wj}\right|
_{\wj=0}=-\left\langle\frac{\wsigma}{g}\right\rangle\Rightarrow\left\langle
\wsigma\right\rangle
=g\mf(g^2,\al)\left\langle\mathcal{O}_{\mathrm{MAG}}\right\rangle\;,
\label{rge60}
\end{eqnarray}
The renormalizability of the action (\ref{rge13bis}) implies that
the action (\ref{rge59bis}) will be renormalizable too. Notice
indeed that both actions are connected through the transformation
\begin{eqnarray}\label{trans1}
    \widetilde{J}&=&\frac{J}{\mathcal{F}(g^2,\al)}\;.
\end{eqnarray}
The tree level off-diagonal masses are now provided by
\begin{eqnarray}\label{rge61bis}
    m_{\mathrm{gluon}}^{\mathrm{off-diag.}}&=&\sqrt{\frac{g\wsigma}{\zeta_0}}\;,\;\;\;\;\;\;\;\;\;\;\;\;
    m_{\mathrm{ghost}}^{\mathrm{off-diag.}}=\sqrt{\al \frac{g\wsigma}{\zeta_0}}\;,
\end{eqnarray}
while the vacuum configuration is determined by solving the gap
equation
\begin{equation}\label{rge62}
    \frac{d\widetilde{V}(\wsigma)}{d\wsigma}=0\;,
\end{equation}
with $\widetilde{V}(\wsigma)$ the effective potential. Minimizing
$\widetilde{V}(\wsigma)$ will lead to a vacuum energy $\Evac(\al)$
which will depend on $\al$ and the hitherto undetermined functions
$f_i(\al)$ \footnote{At first order, $\Evac$ will depend on
$f_0(\al)$, at second order on $f_0(\al)$ and $f_1(\al)$, etc.}.
We will determine those functions $f_i(\al)$ by requiring that
$\Evac(\al)$ is $\al$-independent. More precisely, one has
\begin{equation}\label{rge100}
    \frac{d\Evac}{d\al}=0\Rightarrow\textrm{first order differential equations in $\al$ for }f_i(\al)\;.
\end{equation}
Of course, in order to be able to determine the $f_i(\al)$, we need
an initial value for the vacuum energy $\Evac$. This corresponds to
initial conditions for the $f_i(\al)$. In the case of the linear
gauges,  to fix the initial condition we employed the Landau gauge
\cite{Dudal:2003by}, a choice which would also be possible in case
of the Curci-Ferrari gauges, since the Landau gauge belongs to these
classes of gauges. This choice of the Landau gauge can be motivated
by observing that the integrated operator $\int d^4x A_\mu^A A^{\mu
A}$ has a gauge invariant meaning in the Landau gauge, due to the
transversality condition $\partial_\mu A^{\mu A}=0$, namely
\begin{equation}\label{rge101}
    (VT)^{-1}\min_{U\epsilon SU(N)}\int d^4x\left[\left(A_\mu^A\right)^U \left(A^{\mu A}\right)^U\right]
=\int d^4x (A_\mu^A A^{\mu A}) \;\; \textrm{ in the Landau
gauge}\;,
\end{equation}
with the operator on the left hand side of eq.(\ref{rge101}) being
gauge invariant. Moreover, the Landau gauge is also an all-order
fixed point of the RGE for the gauge parameter in case of the linear
and Curci-Ferrari gauges. At first glance, it could seem that it is
not possible anymore to make use of the Landau gauge as initial
condition in the case of the MAG, since the Landau gauge does not
belong to the class of gauges we are currently considering.
Fortunately, we shall be able to prove that we can use the Landau
gauge as initial condition for the MAG too. This will be the content
of the next section.\\\\Before turning our attention to this  task,
it is worth noticing that, if one would work up to infinite order,
the expressions (\ref{rge53}) and (\ref{rge59bis}) can be
transformed \emph{exactly} into those of (\ref{d2}), respectively
(\ref{rge13bis}) by means of eq.(\ref{trans1}) and its associated
transformation
\begin{equation}\label{trans2}
    \widetilde{\sigma}=\mathcal{F}(g^2,\al)\sigma\;,
\end{equation}
so that the effective potentials $\widetilde{V}(\wsigma)$ and
$V(\sigma)$ are \emph{exactly} the same at infinite order, and as
such will give rise to the same, gauge parameter independent,
vacuum energy.
\section{\label{sec6} Interpolating between the MAG and the Landau gauge.}
In this section we shall introduce a generalized renormalizable
gauge which interpolates between the MAG and the Landau gauge. This
will provide a connection between these two gauges, allowing us to
use the Landau gauge as initial condition. An example of such a
generalized gauge, interpolating between the Landau and the Coulomb
gauge was already presented in \cite{Baulieu:1998kx}. Moreover, we
must realize that in the present case, we must also interpolate
between the composite operator $\frac{1}{2}A_\mu^A A^{\mu A}$ of the
Landau gauge and the gluon-ghost operator
$\mathcal{O}_{\mathrm{MAG}}$ of the MAG. Although this seems to be a
highly complicated assignment, there is an elegant way to treat it.
\\\\Consider again the $SU(N)$ Yang-Mills action with the MAG gauge
fixing (\ref{smn2}). For the residual Abelian gauge freedom, we
impose
\begin{eqnarray}
S_{\mathrm{diag}}^{\prime} &=&\int d^{4}x\left( b^{i}\partial
_{\mu }A^{\mu i}+\overline{c}^{i}\partial
^{2}c^{i}+\occ^i\partial_\mu\left(gf^{iab}A^{\mu
a}c^b\right)+\kappa g f^{iab}A_\mu^a\left(\partial_\mu
c^i\right)\occ^b + \kappa
g^{2}f^{iab}f^{icd}\overline{c}^{a}c^{d}A_{\mu
}^{b}A^{\mu c}\right.  \nonumber \\
&-&\left.\kappa gf^{iab}A_{\mu }^{i}A^{\mu a}(b^{b}-
gf^{jbc}\overline{c}^{c}c^{j})
+\kappa gf^{iab}A^{\mu i}(D_{\mu }^{ac}c^{c})\overline{c}%
^{b}+\kappa g^{2}f^{abi}f^{acd}A_{\mu }^{i}A^{\mu
c}c^{d}\overline{c}^{b}\right) \;, \label{5}
\end{eqnarray}
where $\kappa$ is an additional gauge parameter. The gauge fixing
(\ref{5}) can be rewritten as a BRST exact expression
\begin{eqnarray}
S_{\mathrm{diag}}^{\prime} &=&\int d^{4}x
\left[\left(1-\kappa\right)s\left(\occ^i
\partial_\mu A^{\mu i}\right)+\kappa s\overline{s}\left(\frac{1}{2}A_\mu^i A^{\mu i}\right)\right]\;.\label{6}
\end{eqnarray}
Next, we will introduce the following  generalized mass dimension
two operator,
\begin{equation}\label{7}
    \oo=\frac{1}{2}A_\mu^a A^{\mu a}+\frac{\kappa}{2}A_\mu^i A^{\mu i}+\al \occ^a c^a
    \;,
\end{equation}
by means of
\begin{eqnarray}
S_{\mathrm{LCO}}^\prime &=&s\int d^{4}x\;\left( \lambda \oo +\zeta \frac{\lambda J}{2%
}\right) \;  \label{8} \\
&=&\int d^{4}x\;\left( J\oo +\zeta \frac{J^{2}}{2}-\alpha \lambda
b^{a}c^{a}+\lambda A^{\mu a}D_{\mu }^{ab}c^{b}+ \alpha \lambda
\overline{c}^{a}\left(
gf\,^{abi}c^{b}c^{i}+\frac{g}{2}f\,^{abc}c^{b}c^{c}\right)\right.\nonumber\\&-&\left.\kappa
\lambda c^i\partial_\mu A^{\mu i}+\kappa gf^{iab}\lambda
A_\mu^aA^{\mu i}c^b \vphantom{\frac{\zeta}{2}J^2}\right) \;,
\nonumber
\end{eqnarray}
with $\left( J,\lambda \right)$ a BRST doublet of external
sources,
\begin{equation}
s\lambda =J\;,\;\;\;\;sJ=0\;.  \label{9}
\end{equation}
 As in the case of the gluon-ghost operator
(\ref{ggop}), the generalized operator of eq.(\ref{7}) turns out to
be BRST invariant on-shell.\\\\Let us take a closer look at the
action
\begin{equation}\label{rge200}
    \Sigma^\prime=S_{\mathrm{YM}}+S_{\mathrm{MAG}}+S_{\mathrm{diag}}^\prime+S_{\mathrm{LCO}}^{\prime}+S_{\mathrm{ext}}\;.
\end{equation}
The external source part of  the action, $S_{\mathrm{ext}}$, is the
same as given in eq.(\ref{sexr}).\\\\Also, it can be noticed that,
for $\kappa\rightarrow0$, the generalized local composite operator
$\mathcal{O}$ of eq.(\ref{7}) reduces to the composite operator
$\mathcal{O}_{\mathrm{MAG}}$ of the MAG, while the diagonal gauge
fixing (\ref{6}) reduces to the Abelian Landau gauge (\ref{abgf}).
Said otherwise, for $\kappa\rightarrow0$, the action $\Sigma^\prime$
of eq.(\ref{rge200}) reduces to the one we are actually interested
in and which we have discussed in the previous sections.\\\\
Another special case is $\kappa\rightarrow1$,
$\alpha\rightarrow0$. Then the gauge fixing terms of
$\Sigma^\prime$ are
\begin{equation}\label{rge201}
    S_{\mathrm{MAG}}+S_{\mathrm{diag}}^\prime=\int d^4xs\left(-A_\mu^A\partial^\mu \occ^A\right)=
\int d^4x\left(\occ^A\partial^\mu D_\mu^{AB}c^B+b^A\partial^\mu
A_\mu^A\right)\;,
\end{equation}
which is nothing else than the Landau gauge. At the same time, we
also have
\begin{equation}\label{rge202}
\lim_{(\al,\kappa)\rightarrow(0,1)}\mathcal{O}=\frac{1}{2}A_\mu^A
A^{\mu A}\;,
\end{equation}
which is the  pure gluon mass operator of the Landau gauge
\cite{Verschelde:2001ia,Dudal:2002pq}. \\\\From \cite{Dudal:2002pq},
we already know that the Landau gauge with the inclusion of the
operator $A_\mu^A A^{\mu A}$ is renormalizable to all orders of
perturbation theory. On the other hand, we have already proven the
renormalizability for $\kappa=0$. The complete action
$\Sigma^\prime$, as given in eq.(\ref{rge200}), is BRST invariant
\begin{equation}
s\Sigma^\prime =0\;.  \label{13}
\end{equation}
In \cite{Dudal:2004rx}, we have written down the Ward identities of
this model for $\kappa\neq0$ and general $\alpha$, and we have
proven the renormalizability to all orders of perturbation theory.
It was found that the additional gauge parameter $\kappa$ does not
renormalize in an independent way, while also a generalized version
of the relation (\ref{go}) emerges
\begin{equation}\label{gobis}
    \gamma_\mathcal{O}(g^2)=-2\left(\frac{\beta(g^2)
}{2g^2}-\gamma _{c^{i}}(g^2)\right)\;.
\end{equation}
Summarizing, we have constructed a renormalizable gauge that is
labeled by a couple of parameters $(\al,\kappa)$. It allows us to
introduce a  generalized composite operator $\mathcal{O}$, given by
eq.(\ref{7}), which embodies the  local operator $A_\mu^A A^{\mu A}$
of the Landau gauge as well as the operator
$\mathcal{O}_{\mathrm{MAG}}$ of the MAG. To construct the effective
potential, one sets all sources equal to zero, except $J$, and
introduces unity to remove the $J^2$ terms. A completely analogous
argument as  the one given in section III allows to conclude that
the minimum value of $V(\sigma)$, thus $\Evac$, will be independent
of $\al$ and $\kappa$, essentially because the derivative with
respect to $\al$ as well as with respect to $\kappa$ is BRST exact,
up to terms in the source $J$. This independence of $\al$ and
$\kappa$ is again only assured at infinite order in perturbation
theory, so we can generalize the construction, proposed in section
III, by making the function $\mf$ of eq.(\ref{rge54}) also dependent
on $\kappa$. The foregoing analysis is sufficient to make sure that
we can use the Landau gauge result for $\Evac$ as the initial
condition for the vacuum energy of the MAG. Moreover, we are now
even in the position to answer the question about the existence of a
fixed point of the RGE for the gauge parameter $\al$, which was
necessary to certify that no arbitrary constants would enter the
results for $\Evac$. We already mentioned that the Landau gauge,
i.e. the case $(\al,\kappa)=(0,1)$, is a renormalizable model
\cite{Dudal:2002pq}, i.e. the Landau gauge is stable against
radiative corrections. This can be reexpressed by saying that
$(\al,\kappa)=(0,1)$ is a fixed point of the RGE for the gauge
parameters, and this to all orders of perturbation theory.
\section{Numerical results for $SU(2)$.}
After a quite lengthy formal construction of the LCO formalism in
the case of the MAG, we are now  ready to present explicit
results. In this paper, we will restrict ourselves to the
evaluation of the one-loop effective potential in the case of
$SU(2)$. As renormalization scheme, we adopt the modified minimal
substraction scheme ($\MSbar$). Let us give here, for further use,
the values of the one-loop anomalous dimensions of the relevant
fields and couplings in the case of $SU(2)$. In our conventions,
one has \cite{Shinohara:2001cw,Ellwanger:2002sj,Kondo:2003sw}
\begin{eqnarray}
\label{ex1}    \gamma_{c^i}(g^2)&=&\left(-3-\al\right)\frac{g^2}{16\pi^2}+O(g^4)\;,\\
\label{ex2}
\gamma_\al(g^2)&=&\left(-2\al+\frac{8}{3}-\frac{6}{\al}\right)\frac{g^2}{16\pi^2}+O(g^4)\;,
\end{eqnarray}
while
\begin{equation}\label{ex3}
    \beta(g^2)=-\varepsilon g^2-2\left(\frac{22}{3}\frac{g^4}{16\pi^2}\right)+O(g^6)\;,
\end{equation}
and exploiting the relation (\ref{go})
\begin{equation}
\gamma _{O_{\mathrm{MAG}}}(g^2)=\left( \frac{26}{3}%
-2\alpha \right)\frac{g^2}{16\pi^2}+O(g^4) \;,  \label{ex4}
\end{equation}
a result consistent with that of \cite{Ellwanger:2002sj}. \\\\The
reader will notice that we have given only the 1-loop values of the
anomalous dimensions, despite the fact that we have announced that
one needs $(n+1)$-loop knowledge of the RGE functions to determine
the $n$-loop potential. As we shall see soon, the introduction of
the function $\mathcal{F}(g^2,\al)$ and the use of the Landau gauge
as initial condition allow us to determine the 1-loop results we are
interested in, from the 1-loop RGE functions only.\\\\
Let us first determine the counterterm $\delta\zeta$. For the
generating functional $\mathcal{W}(J)$, we find at 1-loop
\footnote{We will do the transformation of $\mathcal{W}(J)$ to
$\mw(\wj)$ only at the end.}
\begin{eqnarray}\label{ex5}
\mathcal{W}(J)&=&\int
d^dx\left(-\left(\zeta+\delta\zeta\right)\frac{J^2}{2}\right)
+i\ln\det\left[\delta^{ab}\left(\partial^2+\al J\right)\right]
-\frac{i}{2}\ln\det\left[\delta^{ab}\left(\left(\partial^2+J\right)g_{\mu\nu}-\left(1-\frac{1}{\al}\right)
\partial_\mu\partial_\nu\right)\right]\;,\nonumber\\
\end{eqnarray}
and employing
\begin{equation}\label{ex5bis}
   \ln\det\left[\delta^{ab}\left(\left(\partial^2+J\right)g_{\mu\nu}-\left(1-\frac{1}{\al}\right)
\partial_\mu\partial_\nu\right)\right]=\delta^{aa}\left[(d-1)\mathrm{tr}\ln\left(\partial^2+J\right)+\mathrm{tr}\ln\left(\partial^2+\al J\right)\right]\;,
\end{equation}
with
\begin{equation}\label{ex5tris}
\delta^{aa}=N(N-1)=2\textrm{ for }N=2\;,
\end{equation}
one can calculate the divergent part of eq.(\ref{ex5}),
\begin{eqnarray}\label{ex6}
    \mathcal{W}(J)&=&\int d^4x\left[-\delta\zeta\frac{J^2}{2}-\frac{3}{16\pi^2}J^2\frac{1}{\varepsilon}
-\frac{1}{16\pi^2}\al^2
J^2\frac{1}{\varepsilon}+\frac{1}{8\pi^2}\al^2
J^2\frac{1}{\varepsilon}\right]\,.
\end{eqnarray}
Consequently,
\begin{equation}\label{ex7}
    \delta\zeta=\frac{1}{8\pi^2}\left(\al^2-3\right)\frac{1}{\varepsilon}+O(g^2)\;.
\end{equation}
Next, we can compute the RGE function $\delta(g^2,\al)$ from
eq.(\ref{rge7}),  obtaining
\begin{equation}\label{ex8}
    \delta(g^2,\al)=\frac{\al^2-3}{8\pi^2}+O(g^2)\;.
\end{equation}
Having determined this, we are ready to calculate $\zeta_0$. The
differential equation (\ref{rge21a}) is solved by
\begin{equation}\label{ex9}
    \zeta_0(\al)=\al+\left(9-4\al+3\al^2\right)C_0\;,
\end{equation}
with $C_0$ an integration constant. As already explained in the
previous sections, we can  consistently put $C_0=0$. Here, we have
written it explicitly to illustrate that, if $\al$ would
 coincide with the 1-loop fixed point of the RGE for
the gauge parameter, the part proportional to $C_0$ in
eq.(\ref{ex9}) would drop. Indeed, the equations $9-4\al+3\al^2=0$
and $-2\al+\frac{8}{3}-\frac{6}{\al}=0$,  stemming from
eq.(\ref{ex2}), are the same. Moreover, we also notice that this
equation has only complex valued solutions. Therefore, it is even
more important to have made the connection between the MAG and the
Landau gauge by embedding them in a bigger class of gauges, since
then we have the fixed point, even at all orders. In what follows,
it is understood that $\zeta_0=\al$.\\\\
We now have all the ingredients to construct the 1-loop effective
potential $\widetilde{V}_1(\wsigma)$. One obtains
\begin{eqnarray}\label{ex11}
\widetilde{V}_1(\wsigma)&=&\frac{\wsigma^2}{2\zeta_0}\left(1-\left(2f_0+\frac{\zeta_1}{\zeta_0}\right)g^2\right)
+\frac{3}{32\pi^2}\frac{g^2\wsigma^2}{\zeta_0^2}\left(\ln\frac{g\wsigma}{\zeta_0\omu^2}-\frac{5}{6}\right)-
\frac{1}{32\pi^2}\frac{g^2\al^2\wsigma^2}{\zeta_0^2}\left(\ln\frac{g\al\wsigma}{\zeta_0\omu^2}-\frac{3}{2}\right)\;.
\end{eqnarray}
It can be checked explicitly that $\widetilde{V}_1(\wsigma)$ obeys
the renormalization group
\begin{equation}\label{ex12}
    \mu\frac{d}{d\mu}\widetilde{V}_1(\wsigma)=0+\textrm{terms of higher order}\;,
\end{equation}
by using the RGE functions (\ref{ex1})-(\ref{ex4}) and the fact
that the anomalous dimension of $\wsigma$ is given by
\begin{equation}\label{ex13}
\gamma_{\wsigma}(g^2)
=\frac{\beta(g^2)}{2g^2}+\gamma_{\mathcal{O}_{\mathrm{MAG}}}(g^2)
+\mu\frac{\p\ln\mf(g^2,\al)} {\p\mu}\;,
\end{equation}
which is immediately verifiable from eq.(\ref{rge60}).\\\\
We now search for the vacuum configuration by minimizing
$\widetilde{V}_1(\wsigma)$ with respect to $\wsigma$. We will put
$\omu^2=\frac{g\wsigma}{\zeta_0}$ to exclude possibly large
logarithms, and find two solutions of the gap equation $
\left.\frac{d\widetilde{V}_1}{d\sigma}\right|_{\omu^2=\frac{g\wsigma}{\zeta_0}}=0$,
namely
\begin{eqnarray}
    \label{ex14}\wsigma&=&0\;,\\
    \label{ex14bis}y&\equiv&\left.\frac{g^2N}{16\pi^2}\right|_{N=2}=\frac{2\zeta_0}{16\pi^2\left(2f_0\zeta_0+\zeta_1\right)+\al^2\ln\al-\al^2+1}\;.
\end{eqnarray}
The quantity $y$ is the relevant expansion parameter, and should
be sufficiently small to have a sensible expansion.The value for
$\left\langle\wsigma\right\rangle$ corresponding to
eq.(\ref{ex14bis}) can be extracted from the 1-loop coupling
constant
\begin{equation}\label{ex15}
    g^2(\omu)=\frac{1}{\beta_0\ln\frac{\omu^2}{\lms^2}}\;.
\end{equation}
The first solution (\ref{ex14}) corresponds to the usual,
perturbative vacuum ($\Evac=0$), while eq.(\ref{ex14bis}) gives
rise to a dynamically favoured vacuum with energy
\begin{eqnarray}
    \label{ex16}\Evac&=&-~\frac{1}{64\pi^2}\left(3-\al^2\right)
\left(m_{\mathrm{gluon}}^{\mathrm{off-diag}}\right)^4\;,\\
\label{ex16bis}m_{\mathrm{gluon}}^{\mathrm{off-diag}}&=&e^{\frac{3}{22y}}\lms\;.
\end{eqnarray}
From eq.(\ref{ex16}), we notice that at the 1-loop approximation,
$\al^2\leq3$ must be fulfilled in order to have $\Evac\leq0$. In
principle, the unknown function $f_0(\al)$ can be determined by
solving the differential equation
\begin{eqnarray}\label{ex17}
    \frac{d\Evac}{d\al}=0&\Leftrightarrow&2\al\left(m_{\mathrm{gluon}}^{\mathrm{off-diag}}\right)^4
+4\left(\al^2-3\right)\left(m_{\mathrm{gluon}}^{\mathrm{off-diag}}\right)^3\frac{dm_{\mathrm{gluon}}^{\mathrm{off-diag}}}{d\al}=0\nonumber\\
&\Leftrightarrow& \al+\frac{3-\al^2}{y^2}\left(\frac{\p y}{\p
\al}+\frac{\p y}{\p \zeta_0}\frac{\p \zeta_0}{\p \al}+\frac{\p
y}{\p \zeta_1}\frac{\p \zeta_1}{\p \al}+\frac{\p y}{\p
f_0}\frac{\p f_0}{\p \al }\right)=0
\end{eqnarray}
with initial condition $\Evac(\al)=\Evac^{\mathrm{Landau}}$.
However,  to solve eq.(\ref{ex17}) knowledge of $\zeta_1$ is
needed. Since we are not interested in $f_0(\al)$ itself, but
rather in the value  of the vacuum energy $\Evac$, the
off-diagonal mass $m_{\mathrm{gluon}}^{\mathrm{off-diag}}$ and the
expansion parameter $y$,  there is a more direct way to proceed,
without having to solve the eq.(\ref{ex17}).  Let us first give
the Landau gauge value for $\Evac$ in the case $N=2$, which can be
easily obtained from \cite{Verschelde:2001ia,Browne:2003uv},
\begin{equation}\label{ex18}
    \Evac^{\mathrm{Landau}}=-\frac{9}{128\pi^2}e^{\frac{17}{6}}\lms^4\;.
\end{equation}
Since the construction is such that $\Evac(\al)=
\Evac^{\mathrm{Landau}}$, we can equally well solve
\begin{equation}\label{ex19}
-\frac{9}{128\pi^2}e^{\frac{17}{6}}\lms^4=-\frac{1}{64\pi^2}\left(3-\al^2\right)\left(m_{\mathrm{gluon}}^{\mathrm{off-diag}}\right)^4\;,
\end{equation}
which gives the lowest order masses
\begin{equation}\label{ex20}
m_{\mathrm{gluon}}^{\mathrm{off-diag}}=\left(\frac{9}{2}\frac{e^{\frac{17}{6}}}{3-\al^2}\right)^\frac{1}{4}\lms\;,\;\;\;\;\;\;\;\;\;\;\;
m_{\mathrm{ghost}}^{\mathrm{off-diag}}=\sqrt\al\left(\frac{9}{2}\frac{e^{\frac{17}{6}}}{3-\al^2}\right)^\frac{1}{4}\lms\;,
\end{equation}
The result (\ref{ex20}) can be used to determine $y$. From
eq.(\ref{ex16bis}) one easily finds
\begin{equation}\label{ex21}
    y=\frac{36}{187+66\ln\frac{9}{2\left(3-\al^2\right)}}\;.
\end{equation}
We see thus that, for the information we are currently interested
in, we do not need explicit knowledge of $\zeta_1$ and $f_0$. We
want to remark that, if $\zeta_1$ were known, the value for $y$
obtained in eq.(\ref{ex21}) can be used to determine $f_0$ from
eq.(\ref{ex14bis}). This is a nice feature, since the possibly
difficult differential equation (\ref{ex17}) never needs to be
solved in this fashion. Before we come to the conclusions, let us
consider the limit $\al\rightarrow0$, corresponding to the ``real''
MAG $D_\mu^{ab}A^{\mu b}=0$. One finds
\begin{eqnarray}\label{ex22}
    m_{\mathrm{gluon}}^{\mathrm{off-diag}}&=&\left(\frac{3}{2}e^{\frac{17}{6}}\right)^{\frac{1}{4}}\lms\approx2.25\lms\nonumber\;,\\
    y&=&\frac{36}{187+66\ln\frac{3}{2}}\approx0.168\;.
\end{eqnarray}
The relative smallness of $y$ means that our perturbative analysis
should give qualitatively meaningful results.
\section{Discussion and conclusion.}
The aim of this paper was to give analytic evidence, as expressed by
eq.(\ref{ex22}), of the dynamical mass generation for off-diagonal
gluons in Yang-Mills theory quantized in the maximal Abelian gauge.
This mass can be seen as support for the Abelian dominance
\cite{Ezawa:bf,Suzuki:1989gp,Hioki:1991ai} in that gauge. This
result is in qualitative agreement with the lattice version of the
MAG, were such a mass was also reported
\cite{Amemiya:1998jz,Bornyakov:2003ee}. The off-diagonal lattice
gluon propagator could be fitted by $\frac{1}{p^2+m^2}$, which is in
correspondence with the tree level propagator we find. We have been
able to prove the existence of the off-diagonal mass by
investigating the condensation of a mass dimension two operator,
namely $(\frac{1}{2}A_\mu^a A^{\mu a}+\al \occ^a c^a)$. It was shown
how a meaningful, renormalizable effective potential for this local
composite operator can be constructed.  By evaluating this potential
explicitly at 1-loop order in the case of $SU(2)$, the formation of
the condensate is favoured since it lowers the vacuum energy. The
latter does not depend on the choice of the gauge parameter $\al$,
at least if one would work to infinite order in perturbation theory.
We have explained in short how to overcome the problem at finite
order and gave a way to overcome it. Moreover, we have been able to
interpolate between the Landau gauge and the MAG by unifying them in
a larger class of renormalizable gauges. This observation was used
to prove that the vacuum energy of Yang-Mills theory in the MAG due
to its mass dimension two condensate should be the same as the
vacuum energy of Yang-Mills theory in the Landau gauge with the much
explored condensate $\left\langle A_\mu^A A^{\mu A}\right\rangle$.
It is worth noticing that all the gauges, where a dimension two
condensate provides a dynamical gluon mass parameter, such as the
Landau gauge \cite{Verschelde:2001ia}, the Curci-Ferrari gauges
\cite{Dudal:2003gu}, the linear gauges \cite{Dudal:2003by} and the
MAG, can be connected to each other, either directly (e.g.
Landau-MAG) or via the Landau gauge (e.g. MAG and linear gauges).
This also implies that, if $\left\langle A_\mu^A A^{\mu
A}\right\rangle\neq0$ in the Landau gauge, the analogous condensates
in the other gauges cannot vanish either.
\section*{Acknowledgments.}
D.~Dudal would like to thank the organizers of this conference for
the kind invitation to give a talk and for the opportunity to have
many interesting discussions with other participants.

\end{document}